\documentclass[conference]{IEEEtran}
\IEEEoverridecommandlockouts

\usepackage[hidelinks]{hyperref}

\usepackage{amsmath,amssymb,amsfonts}
\usepackage{algorithmic}
\usepackage{graphicx}
\graphicspath{{images/}{../images/}}
\usepackage[font=footnotesize]{caption}
\usepackage[font=footnotesize]{subcaption}
\usepackage{textcomp}
\usepackage{tabularx}
\usepackage{balance}
\usepackage{textcomp}
\usepackage[table,xcdraw]{xcolor}
\def\BibTeX{{\rm B\kern-.05em{\sc i\kern-.025em b}\kern-.08em
    T\kern-.1667em\lower.7ex\hbox{E}\kern-.125emX}}

\definecolor{new_blue}{RGB}{0, 127, 255}

\usepackage{pgfplots}
\pgfplotsset{compat=1.12}
\usepackage{hyperref}
\usetikzlibrary{shapes.multipart,intersections,spy,decorations.fractals,shapes.misc}

\usepackage{pgfplots}

\makeatletter

\usepackage{titlesec}
\usepackage[normalem]{ulem} %to strike the words

\DeclareMathAlphabet\mathbfcal{OMS}{cmsy}{b}{n}

\usepackage{tabularx}
\usepackage{mathfixs}
\usepackage{lipsum}
\usepackage{multirow}
\usepackage{longtable}
\usepackage[nolist]{acronym}
\begin{acronym}[JSONP]\itemsep0pt

\acro{3GPP}{Third Generation Partnership Project}%
\acro{5G}{Fifth Generation}%
\acro{6G}{Sixth Generation}%

\acro{AI}{artificial intelligence}%
\acro{AoA}{angle-of-arrival}%
\acro{AoD}{angle-of-departure}%
\acro{AR}{augmented reality}%
\acro{AP}{access point}%
\acro{ADC}{analog-to-Digital Converter}%
\acro{AE}{antenna Element}%
\acro{AGV}{automatic Guided vehicle}%
\acro{AGV}{automated Guided vehicles}%
\acro{AWGN}{additive white Gaussian noise}
\acro{BER}{Bit Error Rate}%
\acro{BPSK}{Binary Phase-Shift Keying}%
\acro{BRDF}{Bidirectional reflectance Distribution Function}%
\acro{BS}{base station}%
\acro{BF}{Beamforming}%
\acro{BWP}{Bandwidth Parts}
\acro{B5G}{Beyond 5G}

\acro{CA}{Carrier Aggregation}%
\acro{CC}{Component Carrier}
\acro{CDF}{Cumulative Distribution Function}%
\acro{CDM}{Code Division Multiplexing}%
\acro{CDMA}{Code Division Multiple access}%
\acro{CI}{close-in free space reference distance}
\acro{CIR}{Channel Impulse response}
\acro{CPU} {Central Processing Unit}
\acro{CUDA}{Compute Unified Device Architecture}
\acro{CSI}{channel state information}
\acro{C-RNTI}{Cell radio Network Temporary Identifier}
\acro{CRSS}{communication and radar spectrum sharing}
\acro{CFAR}{constant false alarm rate}

\acro{D2D}{Device-to-Device}%
\acro{DL}{downlink}%
\acro{DS}{Delay Spread}%
\acro{DAS}{Distributed antenna System}
\acro{DFRC}{Dual-Functional Radar-Communication}

\acro{EDGE}{Enhanced Data rates for GSM Evolution}%
\acro{EIRP}{Equivalent Isotropic Radiated power}%
\acro{eMBB}{Enhanced Mobile Broadband}%
\acro{eNodeB}{evolved Node B}%
\acro{EoD}{Elevation of Departure}
\acro{ETSI}{European Telecommunications Standards institute}%
\acro{ER}{Effective Roughness}%
\acro{E-UTRA}{Evolved UMTS Terrestrial radio access}%
\acro{E-UTRan}{Evolved UMTS Terrestrial radio access Network}%
\acro{eNB}{enhanced Node B}
\acro{EKF}{Extended Kalman Filter}
\acro{E2E}{End-to-End}
\acro{EoA}{Elevation of Arrival}

\acro{FDD}{frequency Division Duplexing}%
\acro{FDM}{frequency Division Multiplexing}%
\acro{FDMA}{frequency Division Multiple access}%
\acro{FZF}{Full-pilot Zero-Forcing}%
\acro{FoV}{Field of View}%
\acro{FCC}{Federal Communications Commission}%
\acro{FIR}{Finite Impulse response}

\acro{GI}{Global Illumination} %
\acro{GIS}{geographic information System}%
\acro{GO}{geometrical Optics} %
\acro{GPU}{Graphics Processing Unit}%
\acro{GPGPU}{General Purpose Graphics Processing Unit}%
\acro{GPRS}{General Packet radio Service}%
\acro{GSCM}{geometry-based stochastic channel model}
\acro{GSM}{Global System for Mobile Communication}%
\acro{GBSM}{geometry-based stochastic model}%

\acro{H2D}{Human-to-Device}%
\acro{H2H}{Human-to-Human}%
\acro{HD}{High Definition}
\acro{HDRP}{High Definition render Pipeline}
\acro{HSDPA}{High Speed Downlink Packet access}
\acro{HSPA}{High Speed Packet access}%
\acro{HSPA+}{High Speed Packet access Evolution}%
\acro{HSUPA}{High Speed Uplink Packet access}%
\acro{HPBW}{Half-power Beamwidth}%
\acro{HARQ}{Hybrid automatic repeat request}%
\acro{HAD}{hybrid analog-digital}
\acro{HK}{High Knowledge}

\acro{IEEE}{institute of Electrical and Electronic Engineers}%
\acro{inH}{indoor Hotspot} %
\acro{IMT} {international Mobile Telecommunications}%
\acro{IMT-2000}{\ac{IMT} 2000}%
\acro{IMT-2020}{international Mobile Telecommunications 2020}%
\acro{IMT-Advanced}{international Mobile Telecommunications Advanced}%
\acro{IoT}{internet of Things}%
\acro{IP}{internet Protocol}%
\acro{ISAC}{integrated sensing and communications}%
\acro{ITU}{international Telecommunications Union}%
\acro{ITU-R}{\ac{ITU} radiocommunications Sector}%
\acro{IS-95}{interim Standard 95}%
\acro{IRS}{intelligent reflecting Surfaces}%
\acro{IRR}{infrared reflective}%
\acro{i.i.d}{independent and identically distributed}
\acro{IIoT}{industrial internet of Things}%
\acro{industry 4.0}{Fourth industrial revolution}%
\acro{InF}{indoor factory}%
\acro{InF-SL}{InF-Sparse Low}%
\acro{InF-DL}{InF-Dense Low}%
\acro{InF-SH}{InF-Sparse High}%
\acro{InF-DH}{InF-Dense High}%
\acro{InF-HH}{InF-High High}%
\acro{IRR}{infrared reflecting}%
\acro{ISD}{inter-Site Distance}%
\acro{IIoT}{industrial internet of Things}%

\acro{JCAS}{Joint Communication and Sensing}

\acro{KPI}{key performance indicator}%
\acro{KF}{Kalman Filter}
\acro{KLD}{Kullback-Leibler divergence}

\acro{LB} {Light Bounce}
\acro{LIM}{Light intensity Model}%
\acro{LTE}{Long Term Evolution}%
\acro{LTE-Advanced}{\ac{LTE} Advanced}%
\acro{LSCP}{Lean System Control Plane}%
\acro{LSI}{Light Source intensity}
\acro{LoS}{Line of Sight}
\acro{LIDAR}{Light Detection and Ranging}
\acro{LSP}{large scale parameter}
\acro{LK}{Low Knowledge}
\acro{LKF}{Linear Kalman Filter}

\acro{M2M}{Machine-to-Machine}%
\acro{MatSIM}{Multi Agent Transport Simulation}
\acro{METIS}{Mobile and wireless communications Enablers for Twenty-twenty information Society}%
\acro{METIS-II}{Mobile and wireless communications Enablers for Twenty-twenty information Society II}%
\acro{MIMO}{mul\-ti\-ple-in\-put mul\-ti\-ple-out\-put}
\acro{mMIMO}{massive MIMO}%
\acro{mMTC}{massive Machine Type Communications}%
\acro{mmWave}{millimeter wave}%
\acro{MNO}{Mobile Network Operator}
\acro{MU-MIMO}{Multi-user MIMO}
\acro{MMF}{Max-Min Fairness}
\acro{MRT}{Maximum Ratio Transmission}
\acro{MAP-AT}{Map Assisted positioning with angle and Time}
\acro{MPC}{Multipath Component}
\acro{MSE}{Mean Square Error}
\acro{MISO}{multiple input single output}

\acro{NFV}{Network Functions Virtualization}%
\acro{NLoS}{Non Line of Sight}%
\acro{NR}{New radio}%
\acro{NRT}{Non real Time}%
\acro{NYU}{New York University}%
\acro{NGCS}{next generation communication systems}

\acro{O2I}{Outdoor to indoor}%
\acro{O2O}{Outdoor to Outdoor}%
\acro{OFDM}{Orthogonal frequency Division Multiplexing}%
\acro{OFDMA}{Or\-tho\-go\-nal Fre\-quen\-cy Di\-vi\-sion Mul\-ti\-ple access}
\acro{OtoI}{Outdoor to indoor}%

\acro{PDF}{Probability Distribution Function}
\acro{PDP}{power Delay Profile}
\acro{PHY}{Physical}%
\acro{PLE}{Path Loss Exponent}
\acro{PRACH}{Physical Random access Channel}
\acro{PMN}{Predictive Mobility Network}
\acro{PDAF}{Probabilistic Data Association Filter}

\acro{QAM}{Quadrature Amplitude Modulation}%
\acro{QoS}{Quality of Service}%

\acro{RCSP}{receive Signal Code power}
\acro{Ran}{radio access Network}%
\acro{RAT}{radio access Technology}%

\acro{Ran}{radio access Network}%
\acro{RAR}{Random access response}
\acro{RMa}{Rural Macro-cell}%
\acro{RMSE} {Root Mean Square Error}
\acro{RSCP}{receive Signal Code power}%
\acro{RB}{resource Block}
\acro{RT}{ray tracing}%
\acro{Rx}{receiver}%
\acro{RF}{radio frequency}%
\acro{RRU}{remote radio Unit}%
\acro{RIS}{reconfigurable intelligent Surface}%
\acro{RACH}{Random access Channel}%
\acro{RIT}{radio interface Technology}%
\acro{RRC}{radio resource Control}%
\acro{RSSI}{received Signal Strength indicator}
\acro{RCS}{Radar Cross Section}
\acro{RSU}{Road Side Unit}

\acro{SGE}{serious game engineering}%
\acro{SF}{shadow fading}%
\acro{SIMO}{Single input multiple output}%
\acro{SinR}{Signal to interference plus Noise Ratio}
\acro{SISO}{Single input Single Output}%
\acro{SNR}{signal-to-noise ratio}
\acro{SHF}{super high frequency}%
\acro{SSP}{small scale parameter}
\acro{SC}{spatial consistency}
\acro{SA-CSI}{Sensing-assisted \ac{CSI}}
\acro{SAC}{sensing-assisted communications}
\acro{TDD}{Time Division Duplexing}%
\acro{TDM}{Time Division Multiplexing}%
\acro{TD-CDMA}{Time Division Code Division Multiple access}%
\acro{TDMA}{Time Division Multiple access}%
\acro{Tx}{transmitter}
\acro{TRxP}{Transmission and reception point}
\acro{TR}{Technical report}
\acro{ToF}{time of flight}
\acro{Tbps}{terabit-per-second}
\acro{Tb/s}{multi-terabyte per second}

\acro{UE}{user equipment}%
\acro{UL}{uplink}%
\acro{ULA}{uniform linear array}

\acro{V2V}{vehicle-to-vehicle}%
\acro{V2X}{vehicle-to-anything}%
\acro{V2I}{vehicle-to-infrastructure}%
\acro{V2N}{vehicle-to-network}%
\acro{VR}{virtual reality}%

\end{acronym}

\usepackage[ruled,norelsize]{algorithm2e}
\usepackage{listings}

\SetKwSwitch{switch}{case}{other}{switch}{}{case}{other}{end}{end}

\setbox0\hbox{$\xdef\scriptratio{\strip@pt\dimexpr
    \numexpr(\sf@size*65536)/\f@size sp}$}

\newcommand{\scriptveryshortarrow}[1][3pt]{{%
    \vcenter{\hbox{\rule[\scriptratio\dimexpr-.2pt\relax]
               {\scriptratio\dimexpr#1\relax}{\scriptratio\dimexpr.4pt\relax}}}%
   \mkern-4mu\hbox{\let\f@size\sf@size\usefont{U}{lasy}{m}{n}\symbol{41}}}}

\makeatother
\newcommand{\yy}{ \mathbf{y}}

\newcommand{\boldH}{ \mathbf{H} }
\newcommand{\boldF}{ \mathbf{F} }
\newcommand{\boldP}{ \mathbf{P} }
\newcommand{\boldK}{ \mathbf{K} }

\newcommand{\boldI}{ \mathbf{I} }

\newcommand{\xnm}{ x_{n,m} }
\newcommand{\xn}{ x_{n} }

\newcommand{\zz}{ \mathbf{z}}
\newcommand{\hh}{ \mathbf{h}}
\newcommand{\xx}{ \mathbf{x}}

\newcommand{\complexset}[2]{ \mathbb{C}^{#1 \times #2}  }
\newcommand{\complexsett}{ \mathbb{C}  }

\newcommand{\deltaf}{ \Delta f }

\newcommand{\ff}{\mathbf{f}}

\newcommand{\Ntx}{ N_{\rm{T}} }
\newcommand{\Nrx}{ N_{\rm{R}} }

\newcommand{\aaa}{\mathbf{a}}

\newcommand{\atx}{ \aaa_{\rm{T}} }
\newcommand{\arx}{ \aaa_{\rm{R}} }

\newcommand{\ycom}{y^{{\rm{com}}}}
\newcommand{\ysen}{\mathbf{y}^{{\rm{sen}}}}
\newcommand{\hhcom}{\hh^{{\rm{com}}}}

\newcommand{\Ktilde}{\widetilde{K}}

\usepackage{cite}

\begin{document}
\bstctlcite{IEEEexample:BSTcontrol}
\title{Enhancing Sensing-Assisted Communications in Cluttered Indoor Environments through Background Subtraction%ISAC in Dense Indoor Environments via Background Subtraction 
% Sensing-aided Communication in Dense Indoor Environments via Background Subtraction 
% \title{Considering background subtraction for 6G ISAC in Indoor Factory environments\\
\thanks{This work was supported by Hexa-X-II, part of the European Union’s Horizon Europe research and innovation programme under Grant Agreement No 101095759 and the Swedish Research Council (VR grant 2022-03007). The first author would like to acknowledge the support of the Spanish Ministry of Science, Innovation, and University  under the project RTI2018-099880-B-C31.}
}

\author{
Andrea Ramos\IEEEauthorrefmark{1}, Musa Furkan Keskin\IEEEauthorrefmark{2}, 
Henk Wymeersch\IEEEauthorrefmark{2},
Saúl Inca\IEEEauthorrefmark{1},
Jose F. Monserrat\IEEEauthorrefmark{1}
\\
\IEEEauthorrefmark{1}iTEAM Research Institute, Universitat Polit\`ecnica de Val\`encia, Spain\\
\IEEEauthorrefmark{2}Chalmers University of Technology, Sweden}
% \and
% \and
% \and
% \and
% \and

\maketitle

\begin{abstract}
\Ac{ISAC} is poised to be a native technology for the forthcoming \ac{6G} era, with an emphasis on its potential to enhance communications performance through the integration of sensing information, i.e., \ac{SAC}. Nevertheless, existing research on \ac{SAC} has predominantly confined its focus to scenarios characterized by minimal clutter and obstructions, largely neglecting indoor environments, particularly those in industrial settings, where propagation channels involve high clutter density. To address this research gap, background subtraction is proposed on the monostatic sensing echoes, which effectively addresses clutter removal and facilitates detection and tracking of \acp{UE} in cluttered indoor environments with \ac{SAC}. A realistic evaluation of the introduced \ac{SAC} strategy is provided, using \ac{RT} data with the scenario layout following \ac{3GPP} \ac{InF} channel models. Simulation results show that the proposed approach enables precise predictive beamforming largely unaffected by clutter echoes, leading to significant improvements in effective data rate over the existing \ac{SAC} benchmarks and exhibiting performance very close to the ideal case where perfect knowledge of \ac{UE} location is available. 

\end{abstract}

\acresetall
\section{Introduction}
In recent years, wireless communication systems, especially \ac{5G}, has played a crucial role in meeting the diverse and demanding requirements of the Fourth Industrial Revolution, commonly known as Industry 4.0~\cite{5gacia2019}. Industry 4.0 strives to create adaptable and efficient smart factories by integrating applications of \ac{IIoT}. These applications are used in various areas such as logistics, including motion control, smart transportation for inventory management, and collaborative intelligent robots for manufacturing \cite{jiang20213}. These operations frequently occur in indoor scenarios with numerous machines and metallic surfaces, which might pose challenges for wireless connectivity.

\Ac{InF} environments can be more demanding than other indoor deployments~\cite{Ramos2022}. The presence of obstacles results in multiple reflections, leading to a large number of \acp{MPC}. Consequently, establishing high-directional communication may depend on accurately localizing the user's position~\cite{Wen2019, Wymeersch2021}. Moreover, localization can benefit from exploiting characteristics associated with frequency bands~\cite{Lemic2016}, which could be an additional feature using the same \ac{RF} resources.

Within the new era of \ac{6G}, numerous studies exist on high-resolution localization and sensing \cite{de2021,liu2022}. Following this burgeoning interest, novel technologies are poised to emerge, significantly enhancing communication systems through precise localization. One such groundbreaking concept is \ac{ISAC} systems~\cite{Zhang2022}. ISAC is an enabling technology that thrives on the coexistence between sensing and communication capabilities. It is based on the fundamental premise that a sensing system can be seamlessly adapted to multiple radio technologies and diverse environments~\cite{Wymeersch2022}. The convergence of these functionalities promises a host of benefits, including the ability for a communication system to act as a sensor or \ac{SAC}~\cite{6G_predBeam_2023}.

One of the earliest contributions of \ac{SAC} is in ~\cite{liu2020}, where sensing serves communication with novel predictive beamforming (beam tracking). This solution was introduced to mitigate the time overhead caused by downlink and uplink pilots in the conventional \ac{CSI} process. The authors also proposed a simple method to handle beam association of multi-vehicle tracking. A parallel endeavor is evident in \cite{Liu2022a}, which proposes an ID association technique to efficiently predict the state of multiple vehicles. This technique leverages the \ac{KLD} to discern which ID corresponds to each vehicle without needing feedback at every time. Both contributions consider a reflected echo per vehicle for simplicity, an assumption that may not work in cluttered environments. 

Another \ac{SAC} contribution is introduced in \cite{Liu2023}, which proposes dynamic predictive beamforming. This concept involves the adaptation of the beamwidth to track extended vehicle. While this approach might fully illuminate the vehicle as needed, rather than a pencil-sharp beam, the authors leave the scenario geometry out of scope in terms of channel characterization in the system model. In research focused on \ac{ISAC} systems, which depend on a thorough comprehension of the scenario, it is advisable to employ realistic assumptions in channel modeling, such as the inclusion of multipath channels. Although all aforementioned works have contributed to shed light on \ac{SAC} systems, they have all employed an analytical channel model.

\begin{figure}
    \centering
    \includegraphics[width=0.7\columnwidth]{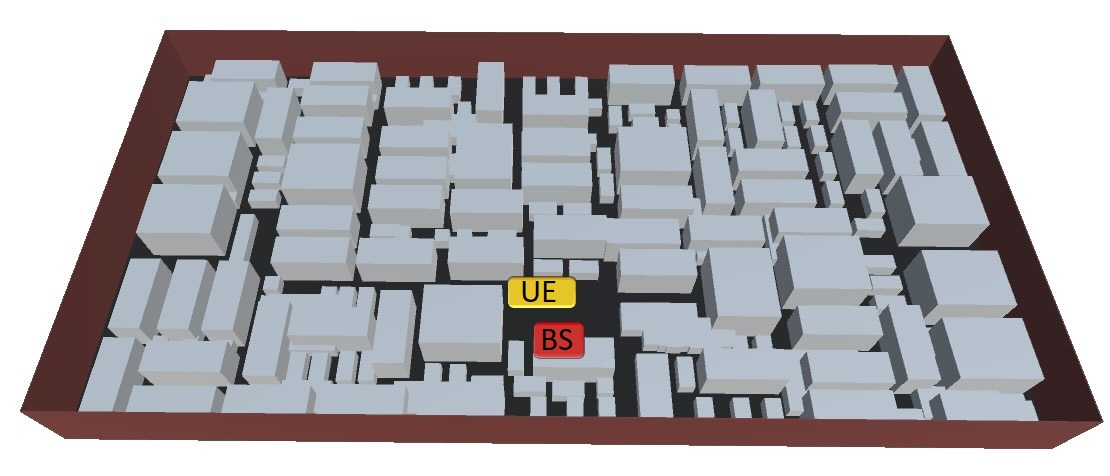}
    \caption{\Ac{InF} sub-scenario with high clutter density (Dense High). The red square represents the \ac{BS}, and the yellow square represents the \ac{UE}.}
    \label{fig:scenario5}\vspace{-5mm}
\end{figure}

% \Ac{SAC} was proposed in ~\cite{liu2020}, where sensing serves communication with novel predictive beamforming constructed by sensing information. This solution was introduced to mitigate the time overhead caused by downlink and uplink pilots in the conventional \ac{CSI} process. However, the authors employed an analytical channel model. In research focused on \ac{ISAC} systems, which rely on a deep understanding of the scenario, it is recommended to utilize realistic channel models. Although the authors of \cite{Liu2023a} present a quantification of the previous solution's overhead reductions, the multi-vehicle beam association assumption (i.e., consider one reflected echo per vehicle) should be studied under multipath propagation effects.
%
In terms of multipath channels, authors in \cite{Liu2023a} model a theoretical channel considering \ac{LoS} and \ac{NLoS} paths, focusing on the quantification of \ac{SAC} overhead reductions. A more elaborate channel modeling is introduced in \cite{cui2023}, which considers \ac{3GPP} \ac{GBSM} for the communication channel and \ac{MIMO} radar multipath channel for sensing one. Even though the latter contribution may be closer to realistic performance, spatial consistency and correlation between sensing and communication channels are inherent features that deterministic channel models (e.g., \ac{RT} dataset) may naturally include. 
Moreover, the previous contributions only focus on \ac{V2X} communication, where environments are "cleaner" with fewer obstructions. Most likely, in more clutter-dense scenarios, detecting the user among obstacles may be challenging, as in \ac{InF}, as shown in Fig.~\ref{fig:scenario5}. In industrial environments, where the reliability of communication is paramount for \ac{IIoT} applications, the clutter in the scenario can directly impact the \ac{LoS} communication. Thus, sensing could play a crucial role in precisely locating the user in such environments.

% Most likely, in more clutter-dense scenarios, detecting the user among obstacles may be challenging, as in \ac{InF}, as shown in Fig.~\ref{fig:scenario5}.

This paper proposes background subtraction as a possible method to facilitate user detection and improve predictive beamforming in \ac{SAC} systems in cluttered indoor environments. This work aims to take a step towards a realistic evaluation of the \ac{ISAC} system by obtaining realistic measurements from a \ac{RT} tool. In order to reveal how beam misalignment can be affected by the number of passive scatterers in the \ac{InF} environment, \ac{SNR} and effective date rate are selected as \ac{KPI}. To establish a comparison, several benchmarks are considered, including \ac{SAC} approaches and communication-only systems.

\section{System model and performance metrics}
% Without loss of generality, we consider an environment comprising. 
% Without loss of generality, it is considered to comprise an environment with $\Ktilde$ scatterers in the communication channel and $K$ echoes in the sensing channel, where reflection coefficients $\{(\alpha^c_k, \alpha^s_k)\}$, \acp{AoD} $\{(\theta^c_k, \theta^s_k)\}$, and delays $\{(\tau^c_k, \tau^s_k)\}$ are the parameters considered to model the received signals.
A downlink communications scenario is considered for an indoor factory environment with a multiple-antenna \ac{BS} and a single-antenna \ac{UE}. The \ac{BS} transmits data symbols to the \ac{UE} while collecting the backscattered signals via a multiple-antenna co-located radar receiver for detecting and tracking the \ac{UE} to enable beam tracking leveraging the principles of \ac{SAC}. %The environment is assumed to include $K$ scatterers (backscattered echoes) within the sensing channel and $\Ktilde$ paths within the communication channel. The parameters of the $k$-th path/scatterer include reflection coefficients ${(\alpha^c_k, \alpha^s_k)}$, \acp{AoD} ${(\theta^c_k, \theta^s_k)}$, and delays ${(\tau^c_k, \tau^s_k)}$.

\subsection{Sensing Signal Model}
The \ac{BS} operates with \ac{MIMO} \ac{ULA}, which has $\Ntx$ transmit antennas and $\Nrx$ receive antennas. Given a precoder $ \ff$, the backscattered signal across $N$ subcarriers can be formulated as:
\begin{align} \label{eq_ynm}
\ysen_{n} = \boldH_{n} \ff \xn + \zz^s_{n} \in \complexset{\Nrx}{1}  ~,
\end{align}
%for $n = 1, \ldots, N$ subcarriers and $m = 1, \ldots, M_T$ symbols, 
where $\xn$ is the \ac{OFDM} transmitted signal {considering a transmitted power $P$}; $\zz^s_{n}$ denotes the \ac{AWGN} with zero mean {with a variance $\sigma^2_N$}, and 
\begin{align}
    \boldH_{n} = \sum_{k=1}^{K} \alpha_k^s e^{-j 2 \pi n \deltaf \tau^s_k} \arx(\theta_k^s) \atx^\top(\theta_k^s)  \in \complexset{\Nrx}{\Ntx} ~,
    \label{eq_hnm}
\end{align}
where $\deltaf$ denotes subcarrier spacing, $\alpha^s_k$ is the complex channel gain, $\theta^s_k$ is the \ac{AoD} (equal to the \ac{AoA}), and $\tau^s_k$ is the delay. The transmit steering vector is the same as the receive steering vector given by the monostatic sensing configuration (i.e., \ac{Tx} and \ac{Rx} are co-located on the same hardware). Thus, both steering vectors, namely $\atx(\theta^s_k)$ and $\arx(\theta^s_k)$ can be denoted as:
\begin{align}
    \aaa(\theta_k^s) = [1, e^{-j \pi \sin(\theta_k^s)}, \dots, e^{-j \pi (N_T-1) \sin(\theta_k^s)}]^\top,
    \label{eq_stV}
\end{align}
considering half-wavelength antenna spacing.

\subsection{Communication Signal Model}
 Assuming a \ac{MISO} downlink communication, the received signal at the \ac{UE} can be written as
\begin{align} \label{eq_ynmcom}
    \ycom_{n} = (\hhcom_{n})^\top \ff \xn + z^c_{n} \in \complexsett  ~,
\end{align}
where
\begin{align}
\label{eq_hcomm}
    \hhcom_{n} = \sum_{k=1}^{\Ktilde} \alpha^c_k e^{-j 2 \pi n \deltaf \tau^c_k}   \atx(\theta^c_k)  \in \complexset{\Ntx}{1}. 
\end{align}
%Here, $\atx(\theta^c_k)$ is the communication transmit array with the same expression as \eqref{eq_stV}. The communication model uses the same transmitted beamforming vector as the sensing model since it is served by sensing information. Thus, the same expression as \eqref{eq_bfpre}. 

\subsection{\Ac{KPI} Selection} 
The goal of this work is to design the communication precoders of the form
\begin{align}
    \ff = \atx^*({\theta}) \in \complexset{\Ntx}{1},
    \label{eq_bfpre}
\end{align}
for a certain beamforming angle ${\theta}$. The selection of the beamforming angle 
depends on the specific method. 
Discrete frames with duration $T$  are considered, comprising beam training and data transmission. After beam training, a vector $\ff$ is determined.
%
% We consider discrete frames of duration $T$, during which beam training and data transmission occur. After beam training, a beamforming vector $\ff$ is determined. 
%
%In the quest for optimal beamforming, the traditional metrics are often \Ac{SNR} and achievable rate. While these metrics provide valuable information, they only tell part of the story. Effective rate, however, might shed light on revealing the true effectiveness and reliability of our beamforming strategies.
%This work determines the effective transmission rate during the frame time denoted by $T'$. However, a certain amount of resources is allocated to the beam training process during the time of $D$. Therefore, to balance and determine the time available for actual data transmission would be the difference between both time resources (i.e., $T_{\text{eff}} = T'-D$).  Thus, to move this assumption into practice, the effective rate shall be calculated as a function of these factors and 
The \ac{SNR} at the \ac{UE} achieved after the beam training period can be written as
  \begin{align}
    \mathrm{SNR}_{r,n} = \frac{P|(\hhcom_{n})^\top \ff |^2}{\sigma^2_N} ~.
 \end{align}
The effective data rate is then formulated as 
\begin{align}
    R_{\text{eff}} = \left(\frac{T-D}{T}\right) \sum^{N}_{n=1} \log_2(1+\mathrm{SNR}_{r,n}),
\end{align}
where $D$ is the time needed for beam training.

\section{Beam Training Methods}

The \ac{UE} moves over discrete time $t$, and at each time $t$, a downlink precoder should be designed. When possible, the time index is removed to lighten the notation. 

\subsection{Conventional Beam Training}
\label{subsec_BT}
Conventional beam training \cite{grythe2015} has been selected as the baseline method (i.e., communication-only system) to compare proposal one. The algorithm aims to find the optimal receiving beamforming weight by looping the best \ac{SNR} among all sampling angle directions. Consider a set of $M$ beamforming directions $\boldsymbol{\Theta}=\{ \theta_1, \ldots, \theta_M\}$ with corresponding precoders $\ff_m=\atx^*({\theta}_m)$. 
%Technically, it may consist of transmitting a symbol $m$ to each angular direction, i.e., $M$ may also represent the number of beam transmissions. 
Considering the channel is static during each time $t$, the corresponding received vector at the \ac{UE} can be denoted as:
\begin{align} \label{eq_covent}
 \ycom_{n,m} = (\hhcom_{n})^\top \ff_m \xnm + z^c_{n,m}.
%\boldH_{t,n,m} \ff_{t,m} x_{t,n,m} + \zz^s_{t,n,m}
\end{align}
%where $\boldH_{t,n,m}$ is generated similarly as \eqref{eq_hcomm} and represent the channel at every time step $t$. 
%The beamforming vector can be analog beamformer represented as  $\ff_{t,m}~=~\aaa_T^*~(\theta_{m})~\in~\complexset{\Ntx}{1}$, for $m=1,\ldots,M$ is the beamforming vector, $x_{t,n,m}$ is the transmitted pilot signal and $\zz^s_{t,n,m}$ is the \ac{AWGN} vector with a variance $\sigma^2_N$.
%Then, the algorithm scans through a range of $M$ angles directions, in this case, $\theta_m \in [-\pi/2 , \pi/2]$. 
%For each direction, $\ycom_{n,m}(\theta_m)$ can give the measured signal. 
The optimal precoder is selected %among many $\theta_m$ 
by maximizing the received power:
\begin{align} \label{eq_BT2}
    \hat{m} &= \arg \max_{m} \sum_{n=1}^N|\ycom_{n,m}|^2,\\
     \hat{\theta}^c &= \theta_{\hat{m}},
\end{align}
which is then used in \eqref{eq_bfpre} for beamforming for communication. 
This process occupies a certain amount of time  $D = M /\Delta f$. Better estimation would be obtained as more time and resources are used to transmit the beams for scanning. However, this process could introduce a high overhead since, in each transmitted frame, the \ac{BS} sends downlink pilots for beam training and then the data transmission.

\subsection{Proposed Method}
\label{sub_subtraction}
% According to the localization literature, \cite{Shahmansoori2018}, obtaining a coarse estimate of the parameter and using it as a starting point for the process is essential. Here is where monostatic sensing can play an important role in assisting communication. However, the estimation can be complex for scenarios with many obstacles, such as \ac{InF}, as the echoes from the user may need to be more distinguishable from the rest of the echoes in the scenario. To this end, we propose a background subtraction method, which will be used as a subroutine in the proposed method.
According to the localization literature \cite{Shahmansoori2018}, obtaining a coarse estimate of the parameter and using it as a starting point for the process is essential.  In this step, monostatic sensing can offer significant assistance without consuming communication resources. Nevertheless, estimating parameters can be complex for scenarios with many obstacles, such as \ac{InF}, as the echoes from the user need to be more distinguishable from the rest of the echoes in the scenario. To this end, the background subtraction is proposed as a subroutine in the proposed method.

\subsubsection{Background Subtraction}
\begin{figure}[!hbt]
    \centering
    \input{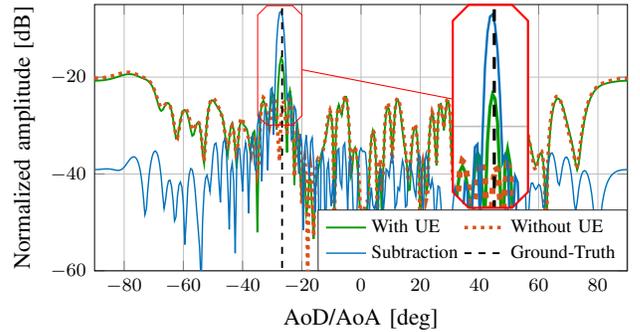}
    \caption{Beam selection estimation based on background subtraction.}
    \label{fig:single_beam_SNR}\vspace{-3mm}
\end{figure}

The estimation of the \ac{AoD} or \ac{AoA} based on background subtraction proceeds as follows:

\begin{itemize}
    \item \textit{Step 1 (Learning stage):} The scenario is previously sensed \emph{without the \ac{UE}}. The received measurement echo signal at the \ac{BS} can be expressed as:
    \begin{align} \label{eq_rr}
        \yy^{\text{ref}}_{n,m} = \Tilde{\boldH}_{n} \mathbf{f}_m \xnm + \zz^s_{n,m} \in \complexset{\Nrx}{1}  ~,
    \end{align}
     where $\Tilde{\boldH}_{n}$ is the sensed channel without considering the \ac{UE}, i.e., considering echoes only from the rest in the scenario. The pairs $(\mathbf{f}_m,\yy^{\text{ref}}_{n,m})$ are stored in a database. The selected beamforming directions cover a fine grid of angles. 
    \item \textit{Step 2 (Inference stage):} Assuming that \ac{BS} transmits with a precoder $\mathbf{f}$, the received signal  can be formulated as in \eqref{eq_ynm}. 
    \begin{itemize}
         \item \textit{Step 2a (background subtraction): } Find the index $m$ for which $\mathbf{f}_m $ is as close as possible to the current precoder $\mathbf{f}$, i.e.,
         \begin{align}
             \hat{m}=\arg \min_m \Vert \mathbf{f}-\mathbf{f}_m\Vert.
         \end{align}
         From the database, the corresponding $\yy^{\text{ref}}_{n,\hat{m}}$ is selected to compute the subtracted signal: 
    \begin{align}
        \label{eq_subtract}
        \yy^{\text{sub}}_{n} = \ysen_{n} - \yy^{\text{ref}}_{n,\hat{m}}.  
    \end{align}
    \item \textit{Step 2b}: Finally, conventional beamforming is applied to estimate \ac{AoD}/\ac{AoA} at monostatic sensing. The best beam to transmit is selected in the angular range $\theta \in [-\pi/2 , \pi/2]$. The best angle is obtained by maximizing the subtracted measurement signal $\yy^{\text{sub}}_{n}$, as
    \begin{align}
        \label{eq_BTsub}
            \hat{\theta}^\text{sub} = \arg \max_{\theta} \sum_{n=1}^N|\arx^H(\theta)\yy_{n}^{\text{sub}}|^2,
    \end{align}
    where $\arx(\theta)$ is the steering vector defined in \eqref{eq_stV}.
    \end{itemize}
    
\end{itemize}
To give a visual evaluation, Fig.~\ref{fig:single_beam_SNR} shows the background subtraction. The optimal angle is obtained as input for the following time step to construct the predictive beamforming.
\begin{algorithm}
\caption{Predictive beamforming} \label{alg_kf} 
\SetAlgoLined
\nl $t=1$, perform conventional beam training to obtain $\hat{\theta}^c$.\\\vspace{-0.7mm}
\nl Set the initial state as: $\hat{\xx}_t = \begin{bmatrix}
\hat{\theta}^c \\ v
\end{bmatrix}$\vspace{-0.7mm}

   \nl \For {$t \in {2,3,...}$}{
        State prediction:\\
        \nl$\hat{\xx}_{t|t-1} = \boldF\hat{\xx}_{t-1}$\\
        
        Predicted state covariance matrix: \\
        \nl$\boldP_{t|t-1} = \boldF\boldP_{t-1}\boldF^\top + \mathbf{Q}_{t-1}$ \\
        
        Kalman gain matrix: \\
        \nl $\boldK_{t} = \boldP_{t|t-1} \mathbf{h}_t (\mathbf{h}^\top_t \boldP_{t|t-1} \mathbf{h}_t + R_t)^{-1}$\\      
        \vspace{0.1cm}
        \nl \switch{}{
        \nl \case{Proposed method}{
        \nl $\mathrm{z}_t = \hat{\theta}^\text{sub}_t$}
        
        \nl \case{Data association method}{
        \nl $\mathrm{z}_t = \arg \min_j | \mathrm{z}_{j,t} - \mathbf{h}_t^\top \hat{\xx}_{t|t-1}|$}
        }
        Innovation (pre-fit residual):\\
        \nl $\mathrm{\Tilde{y}}_t = \mathrm{z}_t - \mathbf{h}_t^\top \hat{\xx}_{t|t-1}$ \\
        Update/correct state: \\
        \nl $\hat{\xx}_t = \hat{\xx}_{t|t-1} + \boldK_{t} \mathrm{\Tilde{y}}_t $ \\
    
        Update covariance matrix: \\
        \nl $\boldP_t = (\boldI - \boldK_t \mathbf{h}_t) + \boldP_{t|t-1}$ 

        The transmit beamforming vector is constructed by:\\
        \nl $ \ff_t = \atx^*({\hat{\theta}_t^\text{pre}}) \in \complexset{\Ntx}{1}$, as \eqref{eq_bfpre} expressed, where $\hat{\theta}_t^\text{pre}$ is the predicted angle obtained from $\hat{\xx}_t$~(line~13).}
\end{algorithm}\vspace{-2mm}

\subsubsection{Predictive Beamforming}
Since the \ac{UE} is moving along a linear trajectory, the \ac{KF} algorithm can be applied to track the dynamic state information. This enables the precise prediction of the state of the \ac{UE}, and this information might be utilized in constructing the predictive beamforming. To delve deeper into the process, it is essential to establish an initial state. In this context, the estimated \ac{AoD} obtained from conventional beam training at $t=1$ (Section \ref{subsec_BT}) is considered the initial state as $\hat{\xx}_t~=~(\hat{\theta}^c, v)$,
% onventional beam training (Section \ref{subsec_BT}) is considered to estimate the \ac{AoD} at $t=1$. Thus, the initial state can be denoted as:
% \begin{align}
%    \hat{\xx}_t = (\hat{\theta}^c, v), \label{eq_ini}
% \end{align}
where $\xx$ is the state vector filtered in every time step, and $v$ is a preset constant velocity. 

The state error covariance matrix should also be determined at the startup. Hence, it can be defined initially as $\mathbf{P} = \mathrm{diag} (\Bar{\sigma}^2_\theta, \Bar{\sigma}^2_v)$, 
%\begin{align}
%    \mathbf{P} = \mathrm{diag} (\Bar{\sigma}^2_\theta, \Bar{\sigma}^2_v)
%\end{align}
which would be naturally updated in every $t$th step. Then, the state vector is predicted using the state evolution model, which in this case is denoted as $\hat{\xx}_{t|t-1} = \boldF\hat{\xx}_{t-1}$, 
%\begin{align}
%  \hat{\xx}_{t|t-1} = %\boldF\hat{\xx}_{t-1}
%\end{align}
where $\boldF$ is the transition matrix defined by the system dynamics where acceleration has been ignored for simplicity. Next, the process noise covariance matrix is related to the uncertainty of the predictive results, which can be expressed as $ \mathbf{Q} = \mathrm{diag} (\sigma^2_\theta, \sigma^2_v)$. 

Following the process of Algorithm \ref{alg_kf}, another important aspect is the uncertainty information, also called measurement. This information obtains the desired state to update the following time steps. Since the uncertainty can be obtained through an independent system \cite{rhudy2017}, this is where the background subtraction can be used. In this explanation, the estimated \ac{AoD} from the background subtraction \eqref{eq_BTsub} at each time step is detonated as $\mathrm{z}_t$, where the noise covariance $R_t$ is a scalar determined by the variance of the ground truth data. 

\subsection{Data Association} \label{sub_asso}
For scenarios where background subtraction is not considered, tracking a single \ac{UE} may be difficult. Given the high clutter density in proposed sub-scenarios, multiple measurements/uncertainties can arise within the area stemming from multiple echoes. In a downlink scenario without uplink feedback from the \ac{UE}, the \ac{BS} might need to associate those \ac{UE} echoes to the predictive state. 

To this end, some extensions of linear \ac{KF} \cite{vo2015} arise in conventional sensing applications. The nearest neighbor method \cite{taunk2019} represents one of the simplest approaches, involving the \ac{BS} in calculating the Euclidean distance between measurements and the predictive state. This method is also outlined in \cite{liu2020} as a beam association technique for multiple targets. Due to its significant presence in the literature, this paper selects it as a baseline method within the \ac{SAC} framework to offer a comprehensive comparison with the proposed approach.

% left fixed width:
\newcolumntype{L}[1]{>{\raggedright\arraybackslash}p{#1}}
 
% center fixed width:
\newcolumntype{C}[1]{>{\centering\arraybackslash}p{#1}}
 
% flush right fixed width:
\newcolumntype{R}[1]{>{\raggedleft\arraybackslash}p{#1}}
\begin{table}
\begin{center}
\begin{tabular}{|L{3cm}|L{4.7cm}|}
\hline
\multicolumn{2}{|c|}{\cellcolor[HTML]{C0C0C0}Scenario layout} \\ \hline    
Sub-scenario                     & Sparse High and Dense High        \\ \hline
Room size (WxL)              & Small-hall $\rightarrow$ L = 120 m, W = 60 m\\ \hline
Ceiling height                  & 10 m                       \\ \hline
BS antenna height            &  8 m              \\ \hline
\ac{UE} width, length, height                    & 0.2 x 0.2 x 0.2 m                       \\ \hline
Clutter density              & \begin{tabular}[c]{@{}c@{}}Low clutter density: 20\%\\ High clutter density 60\%\end{tabular}                               \\ \hline
Clutter height               & \begin{tabular}[c]{@{}c@{}}Low clutter density: 2 m\\ High clutter density: 6 m\end{tabular}                                \\ \hline
Distance between clutter                 & \begin{tabular}[c]{@{}c@{}}Low clutter density: 10 m\\ High clutter density: 2 m\end{tabular}                               \\ \hline
UE Trajectory                & 10 m                 \\ \hline
Time steps               & 100                 \\ \hline
\multicolumn{2}{|c|}{\cellcolor[HTML]{C0C0C0}Configuration parameters} \\ \hline
Carrier frequency             & 28 GHz \\ \hline
Bandwidth                                     & 100 MHz \\ \hline
Total transmit power $P$                      & 21~dBm  \\ \hline
Noise variance $\sigma^2_N$                   & $10^{-9}$ \\ \hline
$N_T$                      & 64 \\ \hline
Time frame $T$                        & 1 ms \\ \hline

\multicolumn{2}{|c|}{\cellcolor[HTML]{C0C0C0} Algorithm 1 assumptions} \\ \hline
$\sigma_{\theta}$ & 1 [deg] \\ \hline
$\sigma_{v}$ & 0.01 [deg/s]\\ \hline
$\Bar{\sigma}_\theta$ & Based on 3dB beam width \\ \hline
$\Bar{\sigma}_v$ & 0.01 [deg/s] \\ \hline
\end{tabular}
\caption{Indoor factory - simulation assumptions.}
\label{tab:tab1}
\end{center}\vspace{-5mm}
\end{table}

The process begins by identifying the most suitable measurement to achieve a reliable predictive state. Therefore, it is assumed that the \ac{BS} is capable of receiving multiple candidate measurements $\mathrm{z}_{j,t}$ from $j = 1, ..., J$ at every $t$. These $J$ candidate measurements are chosen from the backscattered signal with a \ac{CFAR} threshold. In this way, only measurements with similar and highest amplitudes are considered to compute the Euclidean distance between the current a priori prediction and the current observation. In Algorithm \ref{alg_kf}, line 10 defines the aforementioned process as a \emph{case}, in which the closest measurement state would yield the smallest Euclidean distance, as is described in line 11.   

\begin{figure*}
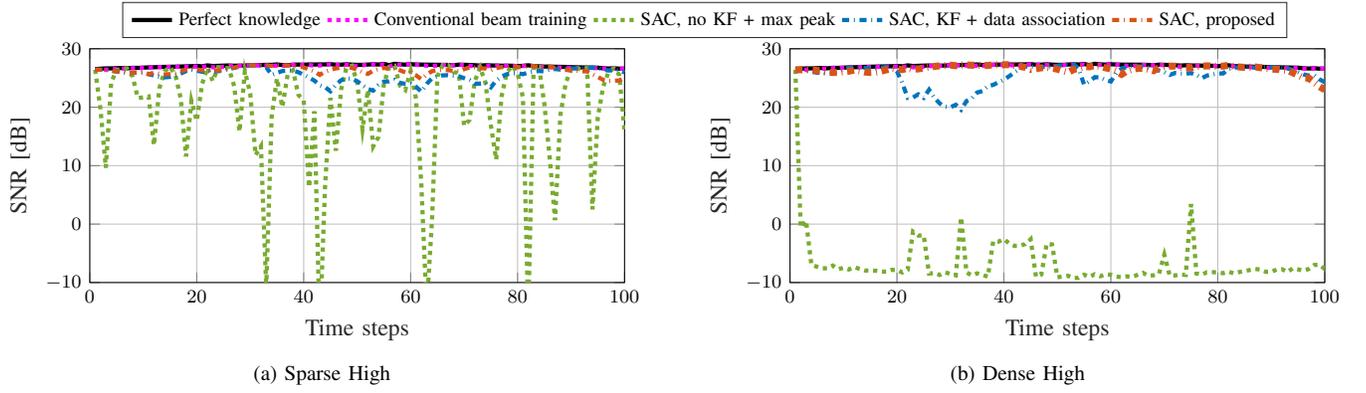

    \centering 
    \begin{subfigure}[b]{0.49\textwidth} 
    \centering
    \input{Figures/tikz/snr_sh1}
    % \vspace{-5mm}
    \caption{Sparse High} \label{fig:snr_sh} 
    \end{subfigure} \hfill 
    \begin{subfigure}[b]{0.49\textwidth} 
    \centering
    \input{Figures/tikz/snr_dh1} 
    % \vspace{-5mm}
    \caption{Dense High} \label{fig:snr_dh} \end{subfigure}
    \caption{\Ac{SNR} performance for the proposed method and benchmark cases.} \label{fig:SNR}
\end{figure*}

\begin{figure*}%[!hbt]
    \centering 
    \begin{subfigure}[b]{0.49\textwidth} 
    \centering
    \input{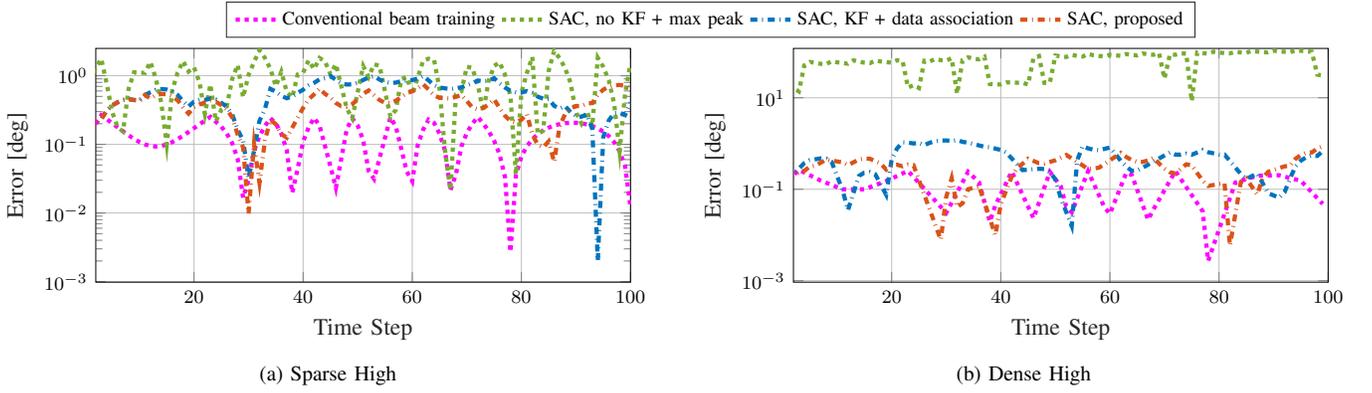}%\vspace{-5mm}
    \caption{Sparse High} \label{fig:error_sh} 
    \end{subfigure} \hfill 
    \begin{subfigure}[b]{0.49\textwidth} 
    \centering
\definecolor{mycolor1}{rgb}{1.00000,0.00000,1.00000}%
\definecolor{mycolor2}{rgb}{0.47000,0.67000,0.19000}%
\definecolor{mycolor3}{rgb}{0.00000,0.45000,0.74000}%
\definecolor{mycolor4}{rgb}{0.85098,0.32549,0.09804}%
\begin{tikzpicture}[scale=1\columnwidth/10cm,font=\footnotesize]
\newcommand\w{8}
\newcommand\h{3.5}
\begin{axis}[%
width=\w cm,
height=\h cm,
at={(0.6in,0.3in)},
scale only axis,
xmin=2,
xmax=100,
xlabel style={font=\color{white!15!black}},
xlabel={Time Step},
ymin=0,
ymax=120,
ymode=log,
ylabel style={font=\color{white!15!black}},
ylabel={Error [deg]},
% axis background/.style={fill=white},
xmajorgrids,
ymajorgrids]
legend columns = 4,
% legend style={at={(1.15,1.2)},anchor=north, legend cell align=left, align=left, draw=white!15!black} ]
% legend pos=north west,
% legend style={legend cell align=left, align=left, draw=white!15!black}
% ]
% legend columns = 2,
% legend style={at={(0.5,-0.25)},anchor=north, legend cell align=left, align=left, draw=white!15!black} ]
% legend style={at={(0.122,0.718)}, anchor=south west, legend cell align=left, align=left, draw=white!15!black}
% ]
\addplot [color=mycolor1, dotted, line width=1.8pt]
  table[row sep=crcr]{%
1	0.156600000000026\\
3	0.265899999999988\\
7	0.154600000000073\\
11	0.0998999999999342\\
15	0.0996999999999844\\
19	0.151499999999942\\
23	0.251600000000053\\
30	0.0305000000000746\\
34	0.240599999999972\\
36	0.142800000000079\\
38	0.0196000000000822\\
39	0.0443000000000211\\
42	0.243600000000015\\
46	0.0233000000000629\\
49	0.239800000000059\\
51	0.115400000000022\\
52	0.0411507999999969\\
53	0.0313550000000049\\
56	0.248158000000004\\
59	0.039985999999999\\
60	0.0290540000000021\\
63	0.229804999999999\\
64	0.205826000000002\\
67	0.0215150000000079\\
68	0.0366280000000074\\
72	0.249970000000005\\
77	0.0331800000000015\\
78	0.00271000000000754\\
83	0.140659999999997\\
87	0.197370000000006\\
91	0.202550000000002\\
95	0.153419999999997\\
99	0.0477900000000062\\
};
% \addlegendentry{Conventional beam training}

\addplot [color=mycolor2, dotted, line width=1.8pt]
  table[row sep=crcr]{%
1	16.3434\\
2	16.3027999999999\\
3	13.7659\\
4	51.2327\\
5	60.7030999999999\\
6	58.177\\
7	55.1546000000001\\
8	54.1357\\
9	63.1202000000001\\
10	59.6083\\
12	55.5948000000001\\
13	66.0931\\
14	61.5948000000001\\
15	58.5997\\
16	57.1079\\
17	65.1193000000001\\
18	62.1339\\
19	60.1514999999999\\
20	58.6721\\
21	66.1957\\
22	63.7221999999999\\
23	16.7516000000001\\
24	18.7837\\
25	16.3185\\
26	64.356\\
27	62.3959999999999\\
28	70.4384\\
29	66.4833\\
30	65.0305000000001\\
31	73.0799\\
32	12.1315\\
33	67.1850999999999\\
34	76.7406\\
35	71.298\\
36	69.3571999999999\\
37	78.918\\
38	19.9803999999999\\
39	19.5443\\
40	19.6095\\
41	21.6759999999999\\
42	21.7436\\
43	21.3122000000001\\
44	21.3818\\
45	19.4521999999999\\
46	74.0233000000001\\
47	81.5951\\
48	21.1672\\
49	21.2398000000001\\
50	83.8124\\
51	80.3846\\
52	77.9588492\\
53	87.031355\\
54	82.103887\\
55	79.676188\\
56	90.748158\\
57	84.81942\\
58	81.890128\\
59	80.460014\\
60	87.029054\\
61	84.597097\\
62	82.664084\\
63	90.229805\\
64	86.794174\\
65	84.357178\\
66	92.918673\\
67	88.978485\\
68	87.036628\\
69	95.09294\\
70	26.14732\\
71	89.19971\\
72	87.24997\\
73	93.29805\\
74	92.34388\\
75	8.38732\\
76	97.92835\\
77	94.46682\\
78	92.50271\\
79	101.53594\\
81	94.59406\\
82	93.61883\\
83	100.64066\\
84	98.15948\\
85	96.67525\\
86	105.68789\\
87	102.19737\\
88	99.70361\\
89	98.20659\\
90	106.70625\\
91	103.70255\\
92	101.19545\\
93	99.18493\\
94	108.67093\\
95	106.65342\\
96	103.63237\\
97	101.60778\\
98	32.07959\\
99	30.54779\\
};
% \addlegendentry{SAC, no KF + max peak}

\addplot [color=mycolor3, dashdotted, line width=1.8pt]
  table[row sep=crcr]{%
1	0.156600000000026\\
2	0.212832089935489\\
4	0.374674728827884\\
5	0.43263920960446\\
7	0.473543959439382\\
8	0.466589613173937\\
9	0.396960776368289\\
11	0.0956165754528229\\
12	0.035147712935796\\
14	0.195362071767846\\
15	0.231132470976689\\
16	0.236787305769383\\
17	0.212477026053406\\
19	0.0757222913760671\\
20	0.4521940178346\\
21	0.781619298669938\\
22	1.06474264584257\\
23	1.04941343711326\\
24	1.00201293670193\\
25	1.03830538179604\\
26	0.923799313111701\\
27	1.04294742379523\\
28	1.12192029862925\\
29	1.16235275929034\\
30	1.16618541816193\\
31	1.1352446720978\\
32	1.19015186353064\\
33	1.08781829594614\\
34	1.04608973716616\\
37	0.848415795942529\\
38	0.830180570326519\\
40	0.723012549386951\\
41	0.63618883937454\\
44	0.262192463869653\\
47	0.280488873298637\\
48	0.276438224354408\\
49	0.237330224655025\\
50	0.165741747512001\\
51	0.0640299093643364\\
53	0.0158270188657639\\
54	0.425694177405802\\
55	0.87504939522367\\
56	0.78289256157295\\
57	0.730506906288468\\
58	0.715590847849057\\
60	0.816880210910938\\
61	0.593272429135311\\
62	0.416888893979177\\
63	0.315091945390293\\
64	0.252667462391727\\
65	0.257635293126455\\
66	0.29514868244307\\
71	0.640470575864242\\
72	0.572730772679961\\
73	0.571053089065373\\
74	0.600683481528137\\
75	0.567869867048429\\
77	0.716607177795709\\
78	0.646147572106884\\
80	0.575389612442123\\
81	0.572454349912988\\
82	0.41586900239912\\
83	0.297937893824255\\
84	0.21605233472269\\
85	0.167788495965013\\
87	0.222224205438494\\
88	0.135841916125841\\
89	0.0823785206175529\\
91	0.0652893015109584\\
92	0.0855861374619451\\
93	0.170054701687022\\
94	0.314538617700563\\
95	0.424041772190208\\
96	0.500943885428342\\
97	0.547451253154989\\
98	0.535170449920884\\
99	0.681595653181773\\
};
% \addlegendentry{SAC, KF + data association}

\addplot [color=mycolor4, dashdotted, line width=1.8pt]
  table[row sep=crcr]{%
1	0.156600000000026\\
3	0.234128066533188\\
4	0.252831456324785\\
5	0.313982584345069\\
6	0.326793600031181\\
8	0.444150865482897\\
11	0.387895127572492\\
12	0.336525696180956\\
14	0.447692645904382\\
15	0.471174140632698\\
16	0.464818347623677\\
18	0.363563131813208\\
19	0.269824737514725\\
20	0.337919190052787\\
21	0.344020255428262\\
22	0.317726797936231\\
23	0.344412656626261\\
24	0.337051660501331\\
29	0.00784364549554084\\
30	0.0677277538509031\\
31	0.173687209024678\\
32	0.040947929341101\\
33	0.0494478018462985\\
34	0.0997329869290553\\
36	0.118314534228759\\
39	0.00991003406255686\\
40	0.0471150838046555\\
42	0.188953884170729\\
43	0.291448800676861\\
44	0.362589614296141\\
45	0.464829047292469\\
47	0.371110010787362\\
48	0.365937989406049\\
50	0.429386558677109\\
52	0.562786130084035\\
55	0.333605153584401\\
56	0.274460860779342\\
58	0.295912224150882\\
62	0.622972146739897\\
63	0.5086003685539\\
64	0.434367596437241\\
65	0.397734560849202\\
66	0.396183848439549\\
67	0.427311831343104\\
68	0.489038102369662\\
69	0.365610799538956\\
70	0.28173618490473\\
71	0.234826800667832\\
72	0.222366967468218\\
74	0.350965681084432\\
75	0.333395329500803\\
76	0.225572880217982\\
77	0.154797958016829\\
78	0.118633758741908\\
81	0.135770329890931\\
82	0.0058430036251309\\
84	0.145440034947214\\
85	0.171635713656713\\
87	0.0770228426109441\\
88	0.175661211064522\\
89	0.240632505069527\\
92	0.316975499733729\\
93	0.326268369770247\\
94	0.461215680736998\\
95	0.561764418472038\\
96	0.630258763908373\\
97	0.638346625803891\\
98	0.773142050580887\\
99	0.874514247016307\\
};
% \addlegendentry{SAC, proposed}

\end{axis}

\end{tikzpicture}%
   % \vspace{-5mm}
    \caption{Dense High} \label{fig:error_dh} \end{subfigure}
    \caption{Error performances to give a visual correlation with the low levels of \ac{SNR}.} \label{fig:error}
\end{figure*}

\begin{figure*}
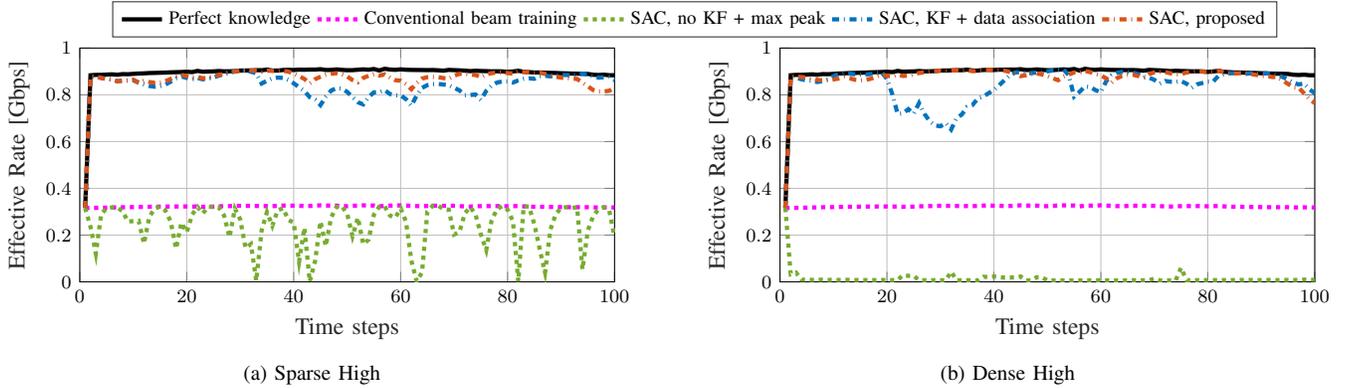
%[!hbt]
    \centering 
    \begin{subfigure}[b]{0.49\textwidth} 
    \centering
    \input{Figures/tikz/rate_sh1} %\vspace{-5mm}
    \caption{Sparse High} \label{fig:rate_sh} 
    \end{subfigure} \hfill 
    \begin{subfigure}[b]{0.49\textwidth} 
    \centering
    \input{Figures/tikz/rate_dh1}
    %\vspace{-5mm}
    \caption{Dense High} \label{fig:rate_dh} \end{subfigure}
    \caption{Effective rate performance for the proposed method and benchmark cases.} \label{fig:rate}
\end{figure*}

\section{Analysis and results}
\subsection{Scenario and Channel Model}
From the \ac{RT} tool, accurate channel parameters are derived from detailed geometric data concerning scatterers and their interactions within the environment. The scenario layout follows the recommendation of the  \ac{3GPP} \ac{InF} channel model in Release 16 \cite{38901_rl16}. This version introduces new channel parameters categorized by their industrial geometric conditions. \ac{InF} channel model incorporates several sub-scenarios classified according to antenna height and clutter density. Specifically, the \emph{Sparse} sub-scenarios signify a low clutter density and are further divided into \emph{Sparse High} and \emph{Sparse Low} relating to high and low \ac{BS} antenna heights. On the other hand, the \emph{Dense} sub-scenario represents a high clutter density, with its divisions based on antenna height as well, namely \emph{Dense High} and \emph{Dense Low}. For this work, \emph{Sparse high} and \emph{Dense high} sub-scenarios are considered since the \ac{UE} detection might be better performed with the antenna \ac{BS} higher than the clutter as a first assumption. Both sub-scenarios are constructed by concrete material, where a metal small \ac{AGV} is considered as a \ac{UE}. The \ac{AGV} follows a straight-line trajectory of 10 meters in front of the \ac{BS} in \ac{LoS}. The acceleration is neglected in these cases since the simulations are snapshots at every time step. From this view, the simulations are performed, and the \ac{RT} channel dataset is extracted. Table \ref{tab:tab1} summarizes the scenario layout and other configuration parameters.

\subsection{Results and Discussion}
In order to provide a comparative analysis, several beam training methods are presented in the following results. \emph{Conventional beam training} is the method explained in Section~\ref{subsec_BT}. \emph{SAC, proposed} is related to Section~\ref{sub_subtraction}, where background subtraction and predictive beamforming is considered.  \emph{SAC, KF + data association} is the benchmark of predictive beamforming, which deals with multiple measurements (Section~\ref{sub_asso}). In addition, \emph{Perfect knowledge} is the ideal case in which the \ac{BS} can perfectly determine the \ac{UE} location, discarding the overhead of downlink and uplink pilots in \ac{SAC} systems. Finally, \emph{SAC, no KF + max peak} determines the optimal beams by finding the angle that maximizes the output of the spatial matched filter applied to the backscattered sensing signal in \eqref{eq_ynm} at every time step without background subtraction (i.e., $\hat{\theta}^\text{sen} =  \arg \max_{\theta} \sum_{n=1}^N|\arx^H(\theta)\ysen_{n}|^2$).The latter method has been selected to illustrate the case of misaligned beamforming without filtering since it estimates the \ac{AoD} without any predictive method.

% However, estimating the \ac{AoD} in this way may result in misaligned beamforming since it also does not incorporate any predictive method.

Fig.~\ref{fig:SNR} shows the level of \ac{SNR} over the time steps for all mentioned performances. In both sub-scenarios, the behavior of \emph{Perfect knowledge} and \emph{Conventional beam training} exhibits some similarities. In conventional communication, beam training aims to establish robust links using highly directional beams that precisely align the transmitter beam with the \ac{UE}. However, this process may introduce high overhead that affects the data transmission. 
On the other hand, it is anticipated that \emph{SAC, no KF + max peak} will yield low \ac{SNR} levels. If the optimal beam is selected by the highest power, it might inadvertently capture echo information from a nearby obstacle, such as the ground. This underscores the importance of incorporating techniques such as predictive beamforming and background subtraction to leverage the sensing information.  
% On the other hand, it is anticipated that \emph{SAC, no KF + max peak} will yield low \ac{SNR} levels, as it involves the backscatter signal to identify the most optimal beam. There is no user feedback in monostatic sensing, making it challenging to discern which echo originates from \ac{UE}. If the highest power selected a codebook, it might inadvertently capture echo information from a nearby obstacle, such as the ground. This underscores the importance of implementing specific processes in sensing, such as data association or the suggested approach of background subtraction.
%
In addition, although the \emph{SAC, KF + data association} obtains a significant level of \ac{SNR}, the \emph{SAC, proposed} slightly overcomes it. This behavior demonstrates that background subtraction may leverage a strong link in \ac{SAC} performance.

Key emphasis is that the \ac{SNR} levels are directly related to detection accuracy. This implies that the accuracy of the information used to direct the transmitted beam should be exceptionally high. Consequently, it is valuable to examine the Error between the ground truth and estimated/predicted information. In Fig.~\ref{fig:error}, there is a correlation between high levels of error and low levels of \ac{SNR} in both sub-scenarios. Even though \emph{Dense High} sub-scenario is the most affected, \emph{SAC, proposed} has lower error levels in more time steps than \emph{SAC, KF + data association}. 

The overhead is another aspect to analyze. Fig.~\ref{fig:rate} shows the effective data rate over time steps. Assuming the overhead of conventional beam training should be considered at the first time step, both \emph{SAC, KF + data association} and the \emph{SAC, proposed} obtain high levels of effective data rate. They neglect the overhead for the rest of the time step using predictive beamforming.

\section{Conclusion}
The evolution of wireless communication systems, transitioning from 5G to the promising realm of 6G, has underscored the significance of precise localization, especially in challenging environments like Indoor factories (InFs). This paper delved into the potential of \ac{ISAC} systems, emphasizing the coexistence of sensing and communication capabilities. Our findings, derived from realistic measurements using a Ray Tracing tool, highlighted the efficacy of background subtraction in enhancing user detection and predictive beamforming in \ac{SAC} systems. The results showcased that while conventional beam training methods can introduce significant overhead, \ac{SAC}, especially with background subtraction, can achieve superior \ac{SNR} levels and effective data rates.

This paper has the potential to pioneer the use of background subtraction in various estimation algorithms under the \ac{ISAC} framework. Specifically, it could significantly impact algorithms tailored for detecting distance, velocity, and positioning. Moreover, delving into real-life scenarios that involve multiple users presents a promising avenue for substantial contributions in forthcoming research endeavors.

%This paper could break new ground for the application of background subtraction in other estimation algorithms within \ac{ISAC} framework. This includes algorithms designed for the detection of distance, velocity, and positioning. Furthermore, exploring real-life scenarios involving multiple users holds significant potential for valuable contributions in future research directions.}

%This is pivotal, as high SNR levels are intrinsically tied to detection accuracy. Furthermore, the correlation between high error peaks and low SNR levels in dense environments accentuates the need for accurate information in directing transmitted beams. In summary, the integration of sensing with communication, bolstered by techniques like background subtraction, holds immense potential for the future of wireless communication, especially in clutter-dense scenarios.
%\section*{Acknowledgment}
\balance 
\bibliography{main}

\end{document}